\documentclass[aps,prr,twocolumn,superscriptaddress]{revtex4-1}
\usepackage{amsmath,bbm,amssymb}
\usepackage{xcolor,graphics,graphicx}
\usepackage[export]{adjustbox}

\usepackage{subfigure}
\usepackage{times}
\usepackage{natbib}
\usepackage{hyperref}
\hypersetup{
colorlinks=true,
linkcolor=blue,
citecolor=blue,
filecolor=magenta,
urlcolor=cyan
}

\usepackage[capitalize]{cleveref}

\newcommand{\be}{\begin{equation}}
\newcommand{\ee}{\end{equation}}
\newcommand{\bea}{\begin{align}}
\newcommand{\eea}{\end{align}}
\newcommand{\id}{\mathbbm{1}}

\newcommand{\sig}{{\boldsymbol{\sigma}}}

\newcommand{\rs}{\hat{\boldsymbol{r}}}
\newcommand{\bmat}{\begin{pmatrix}}
\newcommand{\emat}{\end{pmatrix}}

\newcommand{\x}{\times}

\newcommand{\bh}{\hat{b}}
\newcommand{\ah}{\hat{a}}

\newcommand{\xh}{\hat{x}}
\newcommand{\ph}{\hat{p}}

\begin{document}

\title{Cavity optomechanics assisted by optical coherent feedback}

\author{Alfred Harwood}
\affiliation{Department of Physics \& Astronomy, University College London, Gower Street, WC1E 6BT, London, United Kingdom}
\author{Matteo Brunelli}
\affiliation{Cavendish Laboratory, University of Cambridge, Cambridge CB3 0HE, United Kingdom}
\author{Alessio Serafini}
\affiliation{Department of Physics \& Astronomy, University College London, Gower Street, WC1E 6BT, London, United Kingdom}

\begin{abstract}
We consider a wide family of optical coherent feedback loops acting on an optomechanical system 
operating in the linearized regime. 
We assess the efficacy of such loops in improving key operations, such as cooling, steady-state squeezing and entanglement, as well as optical 
to mechanical state transfer. 
We find that mechanical sideband cooling can be enhanced through passive, interferometric 
coherent feedback, achieving lower steady-state occupancies and considerably speeding up the cooling process; we also quantify the detrimental effect of non-zero delay times on the 
cooling performance.
Steady state entanglement generation in the blue sideband can also be assisted by passive interferometric feedback, which allows one to stabilise 
otherwise unstable systems, though active feedback (including squeezing elements) does not help to this aim.
We show that active feedback loops only allow for the generation of optical, but not mechanical squeezing.
Finally, we prove that passive feedback can assist state transfer at transient times for red-sideband driven systems in the strong coupling regime. 
\end{abstract}

\maketitle

\section{Introduction}
Optomechanical systems, where light modes are coupled to massive mechanical oscillators, have applications in quantum technologies and investigations of fundamental physics~\cite{aspelmeyer2014cavity,OptomechanicsBook}. 
A goal common to both areas is to control coherent properties of mechanical motion, e.g. by cooling to near the ground-state~\cite{Chan11,Teufel11,Peterson16, Qiu2020,Delic20}, generating squeezing~\cite{Pirkkalainen15, Lecocq15} or entanglement between optical and mechanical modes~\cite{Palomaki13,Riedinger16,Gut20}. Since the control is exerted via an optical cavity, a natural question is whether  coherent optomechanical effects can be enhanced by implementing additional operations on the optical field, e.g. by means of feedback loops. A unique possibility in this respect is provided by coherent feedback (CF), where enhanced control is achieved via a measurement-free feedback loop~\cite{Lloyd2000,ZhangReview}. 
In this work, we investigate the efficacy of CF as a way of achieving the goals mentioned above, as well as exploring CF protocols for generating mechanical squeezing and enhancing the transfer of states from the optical to the mechanical mode.
  
\begin{figure}
    \includegraphics[trim={7.5cm 4.5cm 7cm 2cm}, clip, width=0.45\textwidth]{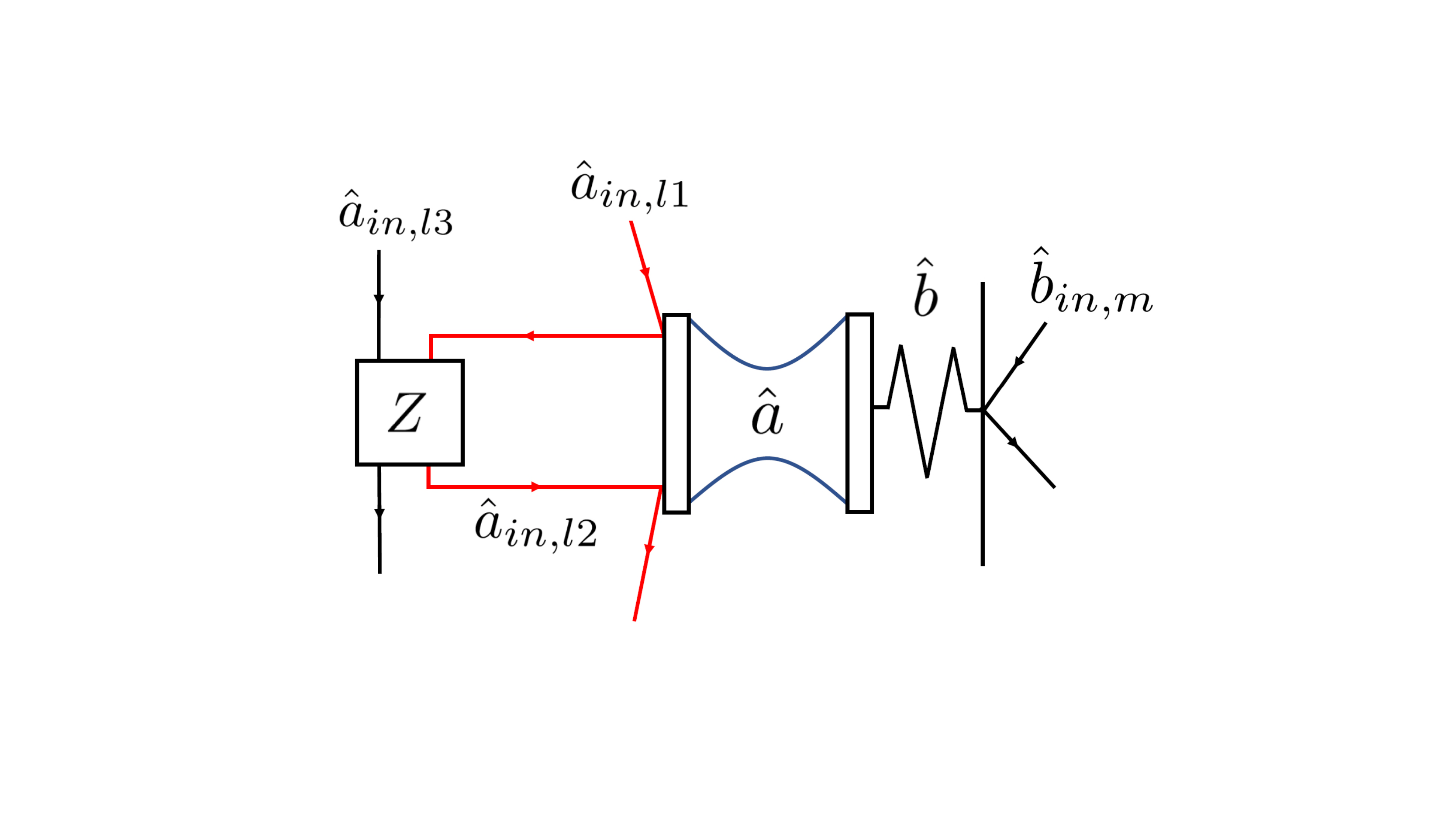} 
    \caption{A schematic diagram of the coherent feedback setup considered. The optical  and mechanical  modes are labelled $\ah$ and $\bh$ respectively. The coherent feedback loop is shown in red,  with the output of interface 1 being subjected to a general operation $Z$, along with ancilla mode $\ah_{in, l3}$ before being fed back into the optical cavity as $\ah_{in, l2}$. Noise on the mechanical mode is indicated by an interaction with an environment given by $\bh_{in, m}$.}    \label{CFDiagram}
\end{figure}

Unlike measurement-based feedback (MF), which involves measurements on the system and uses the results to inform operations, a CF loop is one where quantum information is extracted from a system, processed, and fed back into the system without measurements being performed~\cite{ZhangReview}.
In this study, we consider a CF setup where the light leaking out of one interface of the cavity is allowed to interfere with ancillary modes---the most general process being described by a completely positive (CP) map---before being fed back into the cavity through another interface, see Figure \ref{CFDiagram}. We note, though, that other forms of CF loops are possible, see e.g. \cite{kerckhoff2013}. Quantum feedback through input-output interfaces is often modelled using the SLH and linear quantum feedback network formalisms \cite{combes2017slh, gough2010squeezing, gough2009quantum}. However, here we use the framework for Gaussian coherent feedback developed in \cite{coffee}. 

Cooling the center-of-mass mechanical motion is a key operation and a prerequisite for most quantum protocols \cite{aspelmeyer2014cavity}. Cooling can be implemented in a variety of ways, ranging from parametric feedback cooling (a form of measurement-based feedback)~\cite{ManciniMFcooling1998, VitaliMFCooling2002, Hopkins2003,Vovrosh17,Guo19}, to sympathetic cooling \cite{frimmer2016, hammererSympCooling, camererSympCooling2011,bennett2014coherent} or sideband cooling~\cite{WilsonRae07,Marquardt07}. The latter strategy, in particular, is compatible with the implementation of a CF loop.
A simple question, yet unaddressed so far, is whether CF can be beneficial for mechanical sideband cooling. CF has been investigated in Ref.~\cite{hamerly2012advantages} in relation to cooling, though not specifically for optomechanical setups. The treatment in Ref.~\cite{jacobs2015comparing} 
draws a connection between standard sideband cooling and CF, however without considering the addition of CF loops. In this paper we will use `coherent feedback' to refer to explicit loops constructed using input-output interfaces, as described in the second paragraph. Ref.~\cite{huang2019cooling} considers a setup similar to ours; 
our work generalises this inquiry by obtaining analytic results for a much wider class of feedback loops. We show that, by its ability to tune the effective optical loss rate, CF can considerably reduce the both steady-state mean occupancy and the relaxation time.

Beside cooling, we also consider CF steady-state enhancement of quantum resources, e.g. squeezing and optomechanical entanglement. In optomechanics, entanglement between the mechanical and the optical modes can be generated through blue-detuned sideband drive, which enacts a two-mode squeezing Hamiltonian.
Our analysis shows that interference alone, without involving active operations in the feedback process, is superior for these goals. Active operations, which represent a resource per se, do not appear to be useful in connection with CF. 
CF for the purpose of generating mechanical squeezing in conjunction with mechanical parametric amplification has been investigated in \cite{you2017strong}. In contrast, our study assumes no direct manipulation of the mechanical oscillator.
We also mention that approaches related to CF have been investigated for optomechanical arrays~\cite{Joshi14, vitali2017cf}, and recently delayed CF has been considered for enhancing optomechanical nonlinearity~\cite{Wang17}.

In this paper, we start by introducing the general formalisms of Gaussian states and optomechanical systems in Sections \ref{GaussianSection} and \ref{OptomechanicsSection} respectively. In Section \ref{CFSection} we introduce CF and characterise three different kinds of feedback loops:
passive (interferometric) loops, loops involving squeezing and losses, and loops involving two mode squeezing. Then, in Section \ref{CoolingSection}, we apply these loops to the task of cooling the mechanical oscillator. In the weak coupling regime, we analytically derive the optimal passive CF loop for this purpose before providing numerical evidence that such a setup cannot be outperformed by the active loops considered. We show numerically that passive loops also allow cooling to be enhanced in the strong coupling regime. In the final part of Section \ref{CoolingSection}, we demonstrate that these protocols still perform well when moderate delays are introduced in the loops. In Section \ref{EntanglementSection} we show that, in certain circumstances, CF can be used to stabilise unstable systems in the blue sideband regime, as well as increase the steady state entanglement 
between the optical and the mechanical mode. Again, the passive setups are found to outperform the active loops. In Section \ref{SqueezingSection}, it is shown that active CF can be used to stabilise optical squeezing, but does not lead to steady state mechanical squeezing. Finally, Section \ref{TransferSection} looks at the transfer of a state from the optical to mechanical mode. We find that passive CF can be used to enhance the fidelity of this transfer. 
{}

\section{Gaussian Diffusive Dynamics} \label{GaussianSection}
In what follows, we will use $\ah$ to denote the annihilation operator for the cavity mode of the optomechanical setup and $\bh$ to denote the annihilation operator for the mechanical mode.
We will use $\rs_l= (\hat{x}_l, \hat{p}_l)^{\sf T}$ to refer to the cavity quadrature operators $\hat{x}_l = \frac{1}{\sqrt{2}}(\ah + \ah^\dag)$ and $\hat{p}_l = \frac{i}{\sqrt{2}}(\ah^\dag - \ah)$, and $\rs_m = (\hat{x}_m, \hat{p}_m)^{\sf T}$ to refer to the similarly defined mechanical operators. The notation $\rs = \rs_l \oplus \rs_m$ will be used to denote the total vector of system operators. The operators in $\rs$ must obey the canonical commutation relations (CCR) which are captured by the anti-symmetrized commutator
\be
        [\rs, \rs^{\sf T}] = \rs \rs^{\sf T} - (\rs \rs^{\sf T})^{\sf T} = i\Omega_2 \, ,
\ee
where $\Omega_n$ indicates a $2 n \x 2 n$ matrix of the form
\be
    \Omega_n= \bigoplus_{j=1}^n
    \bmat
        0 & 1 \\
        -1 & 0\\
    \emat \, .
\ee
In the rest of this paper, we will often omit the subscript from $\Omega$, letting the context specify the dimension.
Since we will be restricting our investigation to the Gaussian regime, the state of the system will be entirely characterized by the first and second statistical moments of these operators which are respectively defined as
\be \label{moments}
    \Bar{\boldsymbol{r}}= \text{Tr}[\rho \rs]\quad \mathrm{and} \quad \sig = \text{Tr}[\{(\rs - \Bar{\boldsymbol{r}}), (\rs - \Bar{\boldsymbol{r}})^{\sf T}\} \rho] \, ,
\ee
where $\rho$ is the density operator for the system~\cite{serafini2017quantum}. The covariance matrix $\sig$ is a real, symmetric matrix whose definition involves the symmetrized anti-commutator $\{ \boldsymbol{v}, \boldsymbol{v}^{\sf T} \} =\boldsymbol{v} \boldsymbol{v}^{\sf T} + (\boldsymbol{v} \boldsymbol{v}^{\sf T})^{\sf T}$. In this paper we will not be interested in properties pertaining to the first statistical moments, so will assume $\Bar{\boldsymbol{r}} =0$.

We will model the optomechanical system as being subject to two Hamiltonians, which we will call $\hat{H}_S$ and $\hat{H}_C$. The system Hamiltonian $\hat{H}_S$ involves only system operators and can be written $\hat{H}_S = \frac{1}{2} \rs^{\sf T} H_S \rs$ where $H_S$ is a symmetric matrix. The Hamiltonian $\hat{H}_C$ couples the system to a white noise environment and is written $\hat{H}_C = \rs^{\sf T} C \rs_{in}(t)$ where $C$ is known as the coupling matrix and $\rs_{in}(t)$ is a quantum stochastic process. 

The quantum stochastic process $\rs_{in}(t)$ models the interaction with the environmental continuum of modes as a series of instantaneous interactions with a different mode at each instant \cite{gardiner1985input}. It obeys the white-noise relations
\bea 
    [\rs_{in}(t), \rs_{in}(t')^{\sf T}] &=  i \Omega \delta(t - t') \, , \\ 
    \langle \{\rs_{in}(t), \rs_{in}(t')^{\sf T} \} \rangle &= \sig_{in}\delta(t-t') \, , \label{inputproperties1} \\
    [\rs_{in}(t), \rs_{in}(t)^{\sf T}](dt)^2 &=  i \Omega dt \, , \\
    \quad
    \langle \{\rs_{in}(t), \rs_{in}(t)^{\sf T} \} \rangle (dt)^2 &= \sig_{in} dt \, , \label{inputproperties2}
\end{align}
where $\sig_{in}$ is the covariance matrix of the input states. 

Under such conditions, the Heisenberg evolution of the system modes is given by the stochastic differential equation \cite{serafini2017quantum}
\be
    d \rs(t) = A \rs(t) dt + \Omega C \rs_{in}(t) dt \, ,
\ee
where $A = \Omega H_S + \frac{1}{2}\Omega C \Omega C^{\sf T}$ is known as the drift matrix. In combination with (\ref{inputproperties2}) and (\ref{moments}), this equation can be used to derive the evolution equation for the covariance matrix 
\be \label{driftdiffusion}
    \dot{\sig} = A \sig + \sig A^{\sf T} +D \, ,
\ee 
where $D =\Omega C \sig_{in} C^{\sf T} \Omega^{\sf T}$ is known as the diffusion matrix. If the system is stable, a steady state can be reached where $\dot{\sig}=0$. The condition for this to happen is that the drift matrix must be `Hurwitz',  meaning that the real parts of all its eigenvalues are less than zero.

In this paper we consider a system of one optical and one mechanical mode. The $4 \x 4$ covariance matrix will take the form
\be \label{generalCM}
    \sig = 
    \bmat
        \sig_l & \sig_{lm} \\
        \sig_{lm}^{\sf T} & \sig_m
    \emat \, ,
\ee
where $\sig_{l}$ is the $2\x2$ optical covariance matrix, $\sig_{m}$ is the mechanical covariance matrix and $\sig_{lm}$ captures the correlations between the two modes.

Using this formalism, the covariance matrices of thermal states are proportional to the identity so that $\sig_{th} = N \id$ where $N = 2 \Bar{N} +1$ and $\Bar{N}$ is the mean environmental excitation number. In this paper we will assume that all environmental states are thermal, with a covariance matrix $\sig_{in} = N_l \id_2 \oplus N_m \id_2$ where $N_l$ is the noise on the optical mode and $N_m$ is the noise on the mechanical mode. In cases where the optical mode interacts with two noise fields, we will take them both to be the same temperature, so that $\sig_{in} = N_l \id_4 \oplus N_m \id_2$.
\section{Optomechanical Interactions} \label{OptomechanicsSection}
We consider an optomechanical system where the optical mode, with frequency $\omega_l$ and annihilation operator $\ah$, and the mechanical mode with frequency $\omega_m$ and annihilation operator $\bh$ are radiation-pressure coupled with single-photon coupling strength $g$. The cavity is driven by a laser with frequency $\omega_L$, which couples to the cavity port via a loss rate $\kappa$.
Applying standard linearization technique~\cite{aspelmeyer2014cavity,Paternostro06}, and going into the interaction picture with respect to the free terms $\hat H_0=\omega_l\ah^\dag \ah + \omega_m \bh^\dag \bh$ results in the Hamiltonian
\be \label{Hint}
    \hat{H}_{int}(t) = g(\alpha e^{-i \Delta t}\ah^\dag + \alpha^* e^{i \Delta t}\ah)(e^{-i \omega_m t}\bh + e^{i \omega_m t}\bh^\dag) \, ,
\ee
where we have defined the detuning between the laser and the cavity frequencies as $\Delta = \omega_L - \omega_l$ and $\alpha$ intracavity mean-field amplitude. Without loss of generality, we assume that $\alpha$ is real and define the linearized coupling strength as $G =  \alpha g$. In the regime known as the red sideband, the detuning is set to $\Delta = -\omega_m$. In the `weak coupling regime', the condition $\omega_m\gg G$ holds, and we can make the rotating wave approximation which yields the Hamiltonian
\be \label{red}
    \hat{H}_{red} = G(\ah^\dag \bh + \ah \bh^\dag) \, .
\ee
This Hamiltonian results in the exchange of excitations between the mechanical oscillator and the cavity. Since the environmental noise is typically much lower for the cavity than the mechanical oscillator and the cavity loss rate is typically much higher than the mechanical loss rate, the result of red sideband driving is to cool the mechanical oscillator. 

In the blue sideband regime, the detuning is set to $\Delta = \omega_m$. When the rotating wave approximation is made, the interaction Hamiltonian (\ref{Hint}) becomes
\be \label{blue}
    \hat{H}_{blue} = G(\ah \bh + \ah^\dag \bh ^\dag) \, .    
\ee
This is a two-mode squeezing Hamiltonian, which entangles the mechanical and optical oscillators.

If we wish to operate in the strong coupling regime (ie. with larger $G$ values), we cannot make the rotating wave approximation~\cite{Groblacher09,Teufel11Strong,Brennecke08}. In this case, we retain all terms in Eq.~\eqref{Hint}.
Moving back to the lab frame and expressing the Hamiltonian in terms of the quadrature operators we have
\be \label{HStrongxp}
    \hat{H} = -\frac{\Delta}{2}(\xh_l^2 + \ph_l^2) + \frac{\omega_m}{2}(\xh_m^2 + \ph_m^2) + 2 G \xh_l \xh_m \, .
\ee
The blue and red sideband Hamiltonians are then given by changing the detunings to $\Delta =\omega_m$ and $\Delta = -\omega_m$ respectively.
This Hamiltonian is harder to handle analytically than the simplified versions given in (\ref{red}) and $(\ref{blue})$. We also note that (\ref{HStrongxp}) does not give the fundamental Hamiltonian of the system as, for large enough $G$, it  has a spectrum which is unbounded from below. However, since such values of $G$ are not normally found in experiments, this does not usually pose a problem. 

We will model losses from both the cavity and the  mechanical oscillator using a Hamiltonian corresponding to an exchange of excitations with the environment. This corresponds to a coupling matrix
\be
    C = 
    \bmat
        \sqrt{\kappa}\Omega_1^{\sf T} & 0 \\
        0 & \sqrt{\Gamma_m}\Omega_1^{\sf T}
    \emat \, ,
\ee
where $\kappa$ is the cavity loss rate and $\Gamma_m$ is the mechanical loss rate. 

When dealing with Eqs.~\eqref{red},~\eqref{blue} we will be focusing on the resolved sideband regime $\kappa<\omega_m$, in which the mechanical oscillations are faster than photon dissipation. This is because we are concerned with continuous-wave driving, which becomes less and less effective moving outside this regime. On the other hand, Eq.~\eqref{HStrongxp} holds for any parameter regime.
Finally we note that from now on, all values are given in units where $\omega_m =1$.

\section{Coherent Feedback} \label{CFSection}
We now look at the  effect of adding a CF loop to the cavity mode. A diagram depicting the kind of loop we consider is shown in Fig.~\ref{CFDiagram}. First, we modify the setup so that the optical cavity is now coupled to an environment through two input-output interfaces, each with strength $\kappa$. This  is captured by an environment given by  $\rs_{in}(t) = \rs_{in,l1}(t)\oplus\rs_{in,l2}(t)\oplus\rs_{in,m}(t)$ and a coupling matrix
\be
    C =
        \bmat
            \sqrt{\kappa}\Omega_1^{\sf T} & \sqrt{\kappa}\Omega_1^{\sf T} & 0 \\
            0 & 0 & \sqrt{\Gamma_m}\Omega_1^{\sf T}
        \emat \,,
\ee
where $\rs_{in,l1}$ and $\rs_{in,l2}$ are the environmental modes at the two cavity interfaces and $\rs_{in,m}$ is the mechanical input mode.
A CF loop is achieved by setting $\rs_{in,l2} = E \rs_{out,l1} + F \rs_{in,l3} $, where $\rs_{out,l1}$ is the output mode at interface 1 and $\rs_{in,l3}$ is an ancilla environmental mode taken to be at the same temperature as $\rs_{in, l1}$. The real matrices $E$ and $F$ characterise the CP-map performed on $\rs_{out,l1}$. For a coupling of this kind, the output mode at interface 1 can be written in terms of the input and system modes, using the input-output boundary condition \cite{gardiner1985input, serafini2017quantum}
\be
    \rs_{out, l1}(t) = \sqrt{\kappa} \rs_l(t) - \rs_{in, l1}(t) \, .
\ee
In order for the CP-map to be physical and preserve the CCR, the matrices $E$ and $F$ must satisfy $E\Omega E^{\sf T} + F \Omega F^{\sf T} = \Omega$. 

The effect of such a CF setup is to couple the system to the environment $\rs_{in}(t) = \rs_{in,l1}(t)\oplus\rs_{in,l3}(t)\oplus\rs_{in,m}(t)$ through the modified coupling matrix~ \cite{coffee} 
\be \label{couplingmatrix}
    C_{cf}=
    \bmat
        \sqrt{\kappa} \Omega^{\sf T} -\sqrt{\kappa}\Omega^{\sf T} E & \sqrt{\kappa}\Omega^{\sf T} F & 0\\
        0 & 0 & \sqrt{\Gamma_m}\Omega^{\sf T}
    \emat   \, .
\ee
As we will see, the modified cavity-environment interactions featuring in the two upper blocks  
determine a modification of the dissipation experienced by the system.
CF also introduces extra terms to the system Hamiltonian. In this case, since the feedback loop occurs only on the optical mode, we only get modifications to the optical Hamiltonian, which take the form
\be
    \hat{H}_S \xrightarrow{} \hat{H}_S + \frac{1}{2} \rs_{l}^{\sf T} H_{cf} \rs_l\,,  
\ee
where $H_{cf} = \kappa (\Omega^{\sf T} E+ E^{\sf T}\Omega )$.
These two changes,  to the system-environment coupling and the system Hamiltonian, fully characterise the modifications to the system made by Gaussian CF \cite{coffee}. 

The choice of $E$ and $F$ matrices corresponds to the physical process implemented in the loop. Here, we consider three types of processes which can be implemented in-loop: passive (interferometric) processes, squeezing with losses, and two-mode squeezing. We will now look at the effect of each of these loops, before looking at their applications.

\subsection{Passive Feedback Loops}
First, we consider passive feedback loops. In practice, such loops correspond to interferometric processes involving losses and beam splitters. These processes are referred to as passive as they do not add energy to the mode. Mathematically,  restricting to passive loops means that the matrices $E$ and $F$ satisfy $EE^{\sf T} + FF^{\sf T} = \id$. It turns out that $2\times 2$ real matrices satisfying this property and the CCR take the form
\be \label{passive}
    E =
    \bmat
        a & b \\
        -b & a\\
    \emat\,, \quad
    F =
    \bmat
        c & d \\
        -d & c\\
    \emat \, ,    
\ee
where $a^2 + b^2 + c^2 + d^2 =1$.
We use $C_p$ to denote the modified coupling matrix for passive CF which can be obtained by plugging $E$ and $F$ into (\ref{couplingmatrix}). Under this passive CF setup, the diffusion matrix is given by
\be \label{passivediffusion}
    D_{p} = \Omega C_p \sig_{in} C_p^{\sf T} \Omega^{\sf T}=
    \bmat
 \kappa_{\mathrm{eff}} N_l \id_2 & 0 \\
        0 & \Gamma_m N_m \id_2
    \emat \,,
\ee
where $\kappa_\mathrm{eff} = 2 \kappa(1-a)$. The drift matrix is given by $A_p =  \Omega H_S + \Omega H_p + \frac{1}{2} \Omega C_p \Omega C_p^{\sf T}$ where $H_S$ is the original Hamiltonian matrix, $H_p$ is the Hamiltonian matrix of the modifications due to CF, and $\frac{1}{2} \Omega C_p \Omega C_p^{\sf T}$ captures the diffusive dynamics. While $H_S$ depends on the Hamiltonian being used to drive the cavity, the other two matrices depend only on the specifications of the feedback loop and are given by
\be \label{passivedrift}
    H_p =
    \bmat
        2 \kappa b \id_2 & 0 \\
        0 & 0 \\
    \emat \,,\quad
    \frac{1}{2} \Omega C_p \Omega C_p^{\sf T} = 
    \bmat
        -\frac{\kappa_\mathrm{eff}}{2} \id_2 & 0 \\
        0 & -\frac{\Gamma_m}{2} \id_2
    \emat \, .
\ee
From this, we can see that there are two effects of passive CF: a modification of the optical cavity frequency, and a change in the effective cavity loss rate. Regarding the first feature, a peculiar feature of CF is that the detuning acquires a dependence on the cavity loss rate. The second feature seems especially appealing from an experimental point of view, since cavity loss is usually regarded as a fixed parameter, set by the geometry of the experimental configuration. On the other hand, a CF loop provides a flexible handle to tune cavity decay rate. By tuning the feedback parameters $a$ and $b$, we can exert control over these aspects of the system. In particular, setting $a=1$ means that the output is fed completely back into the cavity, resulting in an effective loss rate $\kappa_\mathrm{eff}=0$. We note that since setting $a=1$ requires a perfect channel with no losses, this case is not feasible in practice. However, since  none of the results in this paper rely the case where $a=1$ this is not a problem. Tuning the feedback loop so that $a=-1$ means that the input at interface 2 interferes constructively with the input at interface 1 and increases the effective loss rate to $\kappa_\mathrm{eff}=4\kappa$. 

\subsection{Loops containing squeezing and losses} \label{SqueezingCF}
It is worthwhile to extend our treatment to encompass feedback loops where squeezing operations are allowed. 
This choice is certainly relevant in light of the numerous applications of squeezed light to improving the operation of optomechanical systems~\cite{Huang09,Huang09Normal,Jahne09,Clark17,Asjad16,Lu15}.  
The second type of loop we consider is thus one where the feedback mode is subject to losses followed by squeezing. Such feedback is no longer passive as the action of squeezing adds energy. This setup is modelled with $E$ and $F$ matrices given by
\be \label{Sq}
E_z = 
    \bmat
        \eta z & 0 \\
        0 & \frac{\eta}{z}
    \emat
    \quad \text{and} \quad
    F_z =
    \bmat
        \sqrt{1-\eta^2} z & 0 \\
        0 & \frac{\sqrt{1-\eta^2}}{z}
    \emat \, ,
\ee
where $z>0$ is the squeezing parameter $0<\eta<1$ parameterises the losses. When $\eta=0$, the feedback mode is entirely replaced by the noise mode before being squeezed, and when $\eta=1$, the feedback mode is not subject to any losses before being squeezed.

By plugging these matrices into (\ref{couplingmatrix}) we obtain the effective coupling matrix for this setup, which we will call $C_z$. The resulting diffusion matrix, $D_{z} = \Omega C_z \sig_{in} C_z^{\sf T} \Omega^{\sf T}$, is diagonal with entries 
        $\kappa N_l (1 - 2 \eta z + z^2)$, $\kappa N_l (1- \frac{2\eta}{z} + \frac{1}{z^2} )$, $\Gamma_m N_m$ and $\Gamma_m N_m$.        
The drift matrix for this setup instead takes the form $A_z =  \Omega H_S + \Omega H_z + \frac{1}{2} \Omega C_z \Omega C_z^{\sf T}$ 
where, again, $H_z$ captures modifications to the Hamiltonian from CF and $\frac{1}{2} \Omega C_z \Omega C_z^{\sf T}$ 
captures the diffusive dynamics. These are given by
\bea
    H_z&= 
    \bmat
        \sigma_x \kappa \eta ( z - \frac{1}{z}) & 0 \\
        0 & 0 \\
    \emat \,,\quad  \\
    \frac{1}{2}\Omega C_z\Omega C_z^{\sf T} &= 
    \bmat
        \frac{1}{2}\kappa(\frac{\eta}{z} + \eta z -2) \id_2 & 0 \\
        0 & -\frac{\Gamma_m}{2}\id_2
    \emat \, ,
\end{align}
where $\sigma_x$ is the Pauli $x$-matrix. We can see that this kind of feedback results in  modifications to the diffusive dynamics, as well as the addition of a squeezing Hamiltonian to the light mode.

\subsection{Loops containing two-mode squeezing} \label{TMSsection}
The third type of loop we consider is one where the feedback mode, along with the ancilla, is subject to a two-mode squeezing operation before the ancilla is traced out. First we consider loops of this kind. Then, we look at loops containing phase shifters as well as two mode squeezing.

The setup without phase shifters is modelled using $E$ and $F$ matrices given by
\be \label{TMS}
    E_{T} = \cosh{r} \id_2 \quad F_{T} = \sinh{r} \sigma_z \, ,
\ee
where $r$ is the two-mode squeezing parameter, and $\sigma_z$ is the Pauli $z$-matrix.
Again, we can obtain $C_{T}$, the effective coupling matrix  for this system, by plugging these matrices into (\ref{couplingmatrix}). From this we can find the diffusion matrix
\be
    D_{T} = 
    \bmat
        \kappa N_l (2 \cosh^2{r} - 2 \cosh{r}) \id_2 & 0 \\
         0 & \Gamma_m N_m \id_2 \\
    \emat \, .
\ee
For this setup, there are no modifications to the system Hamiltonian as $\Omega^{\sf T} E_{T} +  E_{T}^{\sf T}\Omega = 0$. This means that we can write the drift matrix as $A_{T} = \Omega H_S + \frac{1}{2}\Omega C_{T} \Omega C_{T}^{\sf T}$ with 
\be
    \frac{1}{2}\Omega C_{T} \Omega C_{T}^{\sf T} = 
    \bmat
        \kappa (\cosh{r} -1)\id_2 & 0 \\
        0 & -\frac{\Gamma_m}{2}\id_2
    \emat \, .
\ee
We note that, since $\cosh{r}\geq 1$, this kind of CF will always destabilise optomechanical setups by increasing the eigenvalues of the drift matrix. 

Since we are often interested in steady states,  which are not achievable when the drift matrix has postive eigenvalues,  we might ask if there is a setup involving two mode squeezing which does not destabilise the system. Let us define new $E$ and $F$ matrices $ E_{S} =-E_T \quad F_{S} =-F_T $,
which lead to a diffusion matrix
\be
    D_{S} = 
    \bmat
        \kappa N_l (2 \cosh^2{r} + 2 \cosh{r}) \id_2 & 0 \\
        0 & \Gamma_m N_m \id_2
    \emat \, .
\ee
As before, there are no modifications to the Hamiltonian matrix, but the diffusive dynamics are characterized by the matrix
\be
    \frac{1}{2} \Omega C_{S} \Omega C_{S}^{\sf T} = 
    \bmat
        -\kappa(1+\cosh{r})\id_2 & 0 \\
        0 & \frac{\Gamma_m}{2}\id_2
    \emat \, .
\ee
Notice that, if we define $\kappa_{S} = 2\kappa(1+\cosh{r})$ and $N_{S} = N_l \cosh{r}$ then we can write
\bea \label{TMSdriftdiffusion}
    D_S&=
    \bmat
        \kappa_{S} N_S \id_2 & 0 \\
        0 & \Gamma_m N_m \id_2
    \emat \; ,
    \\
    \frac{1}{2} \Omega C_{S} \Omega C_{S}^{\sf T} &=
    \bmat
        -\frac{\kappa_S}{2}\id_2 & 0 \\
        0 & -\frac{\Gamma_m}{2}\id_2
    \emat \; .
\end{align}
In other words, these loops can be characterized by modifications to the cavity loss rate and the temperature of the optical environment. This means that such loops may allow systems to be stabilized (by increasing $\cosh{r}$ and therefore $\kappa_S$) at the cost of increasing the noise on the optical mode (due to the accompanying increase in $N_S$).

\section{coherent feedback-enhanced sideband cooling} \label{CoolingSection}
Sideband cooling of the mechanical oscillator is achieved by driving the cavity with a detuning $\Delta = -\omega_m$~\cite{WilsonRae07,Marquardt07}. In this section we look at passive feedback loops and  loops containing squeezing as characterized by the $E$ and $F$ matrices given in (\ref{passive}) and (\ref{Sq}) respectively. We do not investigate loops involving two mode squeezing. The reason for this is that loops characterized by $E$ and $F$ matrices (\ref{TMS}) lead to non-Hurwitz drift matrices, and therfore do not reach a steady state. Loops generated by $E_{S},\,F_{S}$,  do lead to a steady state, but only allow modifications to the effective cavity loss rate at the expense of higher noise on the optical mode. Since passive feedback allows for these modifications without the extra noise, we can state that, for a given effective loss rate, passive feedback will outperform the two-mode squeezing feedback. For this reason, we do not consider the stable CF loops with two mode squeezing.

The efficacy of a cooling protocol will be determined by the steady-state entropy of the mechanical mode. To do this, we use the fact that a covariance matrix for a single mode Gaussian state can always be written in the form $\sig =  \nu S S^{\sf T}$ where $S$ is a symplectic matrix which satisfies $S\Omega S ^{ \sf T} = \Omega$ and $\nu$ is known as the symplectic eigenvalue of $\sig$. Single-mode Gaussian states have the convenient property that all entropies of the state are increasing functions of $\nu$ \cite{serafini2017quantum}. We also note that, since $\text{Det}S = 1$, we can write $\nu = \sqrt{\text{Det}\sig} 
= 1/{\rm Tr}[\varrho^2]$, where $\varrho$ is the Gaussian quantum state under examination. 
Note also that when the states involved are thermal states, with $\sig \propto \id_2$, the symplectic eigenvalue and the regular eigenvalue of the state coincide. In plots, we will use the average mechanical excitation number, which is given by $\Bar{N}= \frac{\nu -1}{2}$.

\subsection{Passive Feedback}
\subsubsection{Weak Coupling}
\begin{figure*}
    \centering
    \subfigure[]{\includegraphics[width=0.3\textwidth]{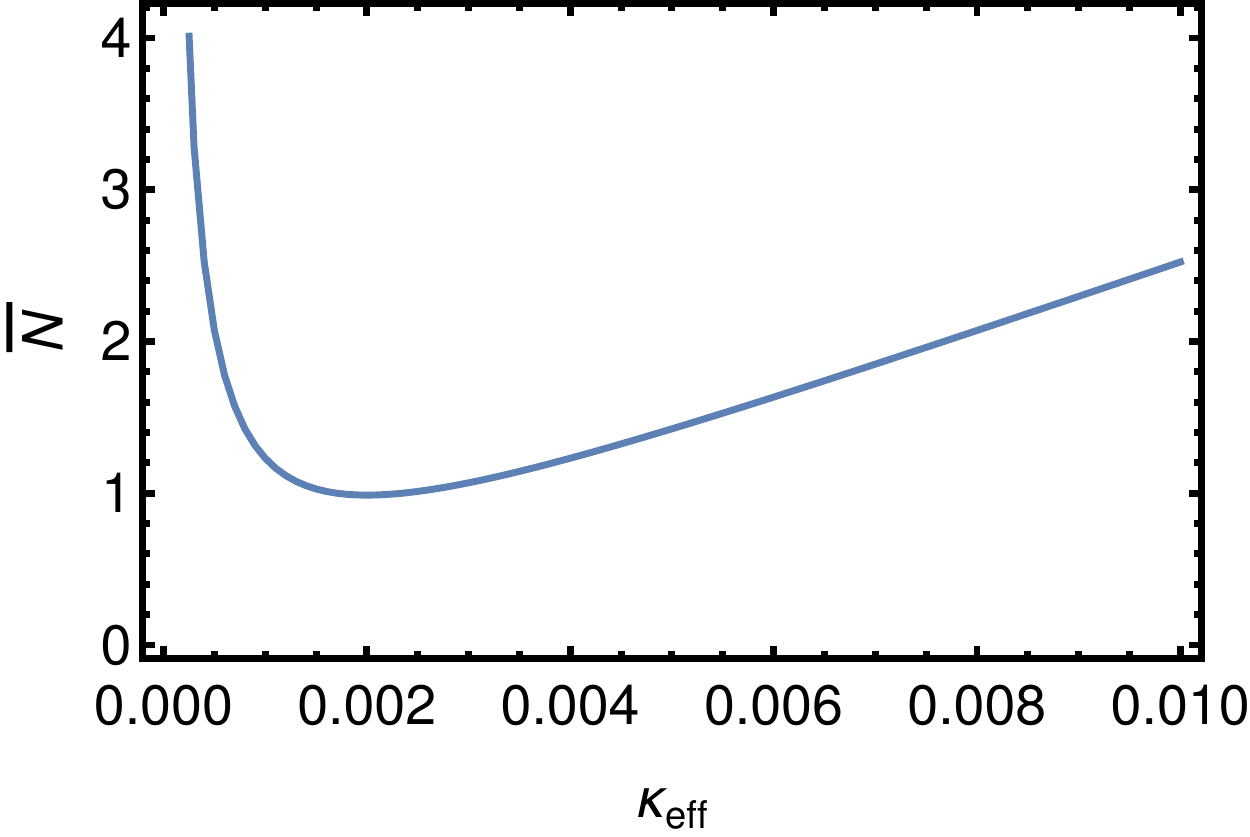} \label{sigmvskeff}} 
    \hfill
    \subfigure[]{\includegraphics[width=0.3\textwidth]{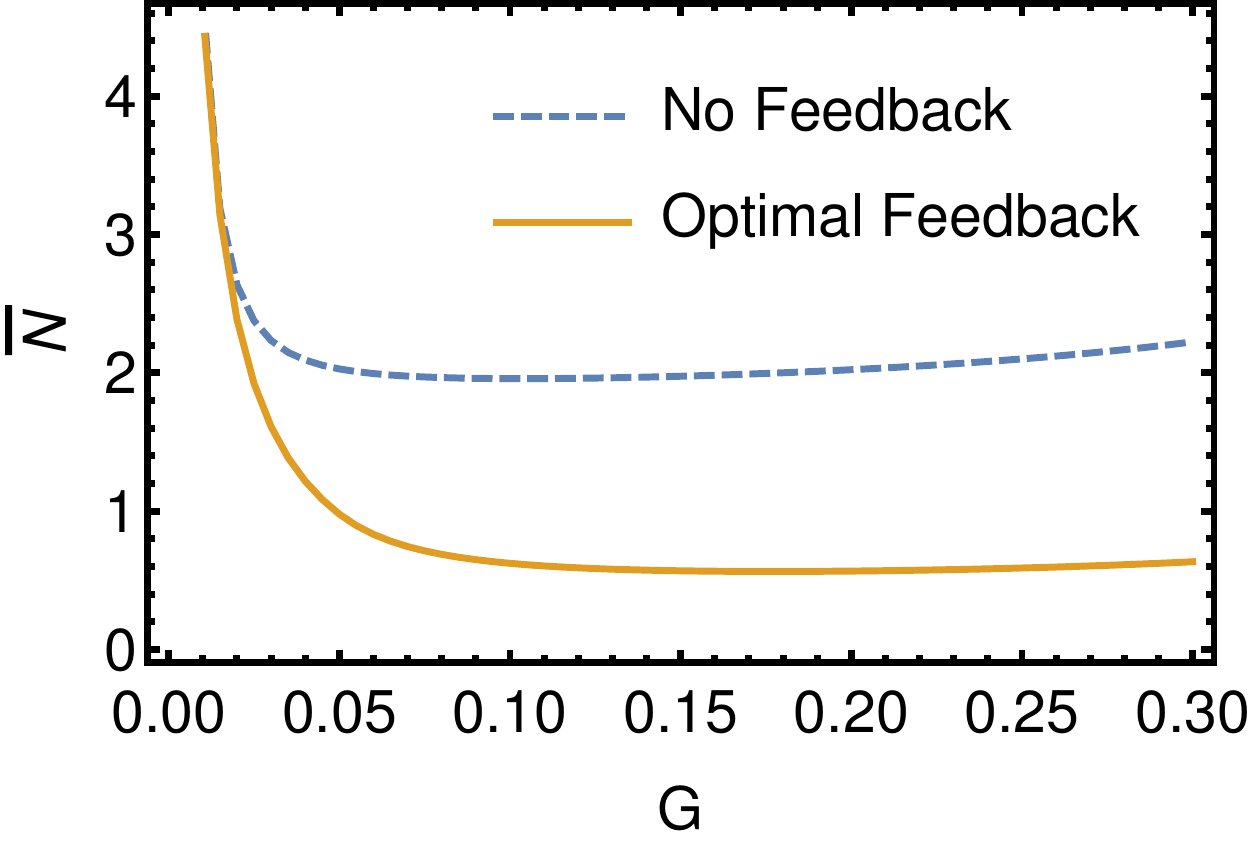} \label{StrongCouplingCooling}} 
    \hfill
    \subfigure[]{\includegraphics[width=0.32\textwidth]{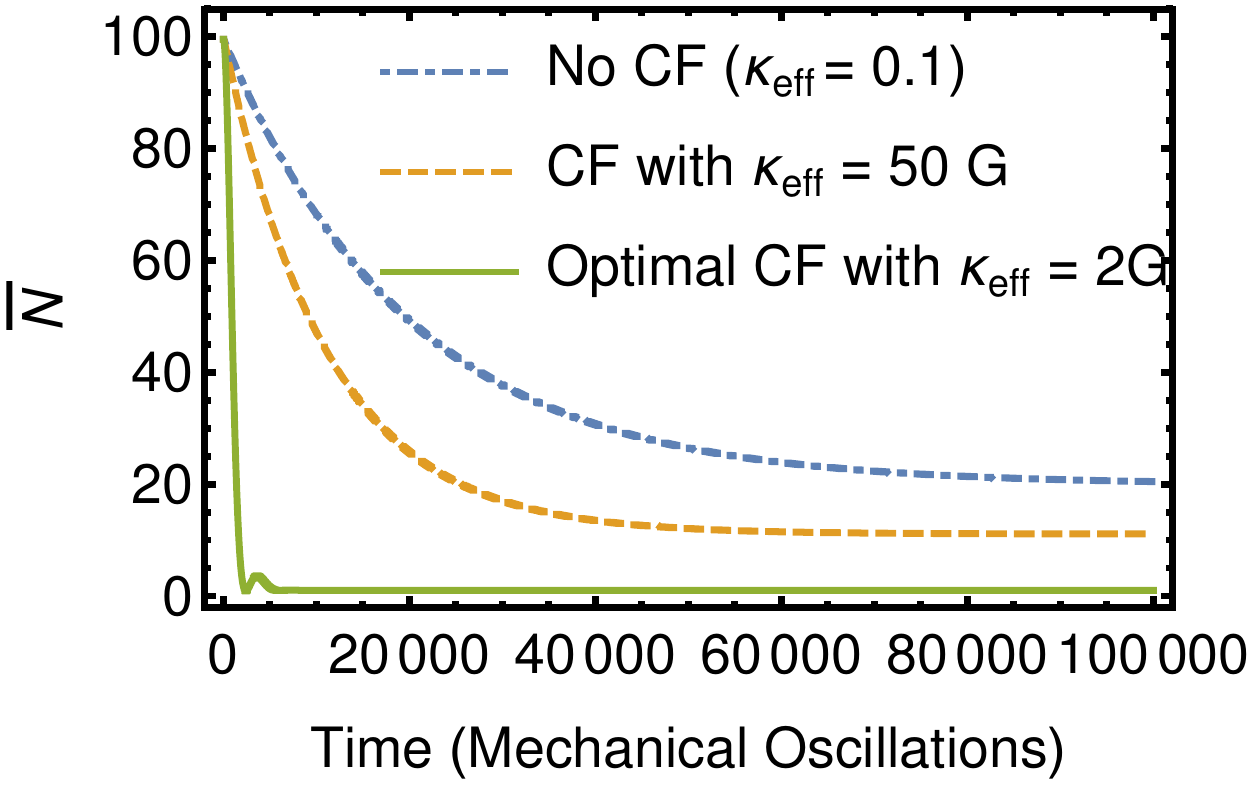} \label{TransientDyanmics}}
    \caption{(a) A plot of the average steady state mechanical excitations against $\kappa_\mathrm{eff}$ (in units of mechanical frequency) for a system operating in the weak coupling, red-sideband regime. The parameters used are $\kappa = 0.1$, $\Gamma_m = 10^{-5}$, $G=10^{-3}$, $N_l=1$, $ N_m =200$, in units where the mechanical frequency is equal to 1. The mechanical temperature is minimized at $\kappa_\mathrm{eff} =2G=2\x 10^{-3}$, but any modification which moves $\kappa_\mathrm{eff}$ towards this optimal value improves the steady state cooling. (b) A plot of the steady state mechanical excitations $\overline{N}$ against coupling strength $G$ (in units of mechanical frequency) for a system in the strong coupling red sideband regime with $\kappa=0.025$, $\Gamma_m = 10^{-3}$,  $N_l=1$, $N_m = 100$. The blue dashed line indicates the steady state cooling when no feedback is used and the orange solid indicates the steady state cooling achievable when 
    passive coherent feedback is numerically optimized for the coupling strength. (c) A plot of the average mechanical excitations against time for the setups where $\kappa_\mathrm{eff} = 0.1$ (no feedback), $\kappa_\mathrm{eff} = 50G= 5\x10^{-2}$ and $\kappa_\mathrm{eff} = 2\x10^{-3} = 2G$ with $\Gamma_m = 10^{-5}$, $G=10^{-3}$, $N_l=1$, $ N_m =200$.}
\end{figure*}

Sideband cooling in the weak coupling regime means that the system is subject to Hamiltonian (\ref{red}). Combining this with the Hamiltonian modifications and diffusive dynamics from (\ref{passivedrift}) leads the system to evolve according to the drift matrix given by 
\be
    A = 
    \bmat
     -\frac{\kappa_\mathrm{eff}}{2} & 2 \kappa b  & 0 & G\\
        -2 \kappa b & -\frac{\kappa_\mathrm{eff}}{2} & -G & 0 \\
        0 & G & -\frac{\Gamma_m}{2} & 0\\
        -G & 0 & 0 & -\frac{\Gamma_m}{2}
    \emat \, ,
\ee
and a diffusion matrix given by (\ref{passivediffusion}). 
Under this setup, the steady state mechanical covariance matrix is a thermal state $\sig_m = \sigma_m \id_2$ with eigenvalue
\begin{widetext}
\be \label{sigcf}
    \sigma_{m} = \frac{\Gamma_m \kappa_\mathrm{eff} (16 b^2 \kappa^2 + (\Gamma_m + \kappa_\mathrm{eff})^2) N_m + 4 G^2 (\Gamma_m + \kappa_\mathrm{eff}) (\kappa_\mathrm{eff} N_l + \Gamma_m N_m)}{4 G^2 (\Gamma_m + \kappa_\mathrm{eff})^2 + \Gamma_m \kappa_\mathrm{eff} (16 b^2 \kappa^2 + (\Gamma_m + \kappa_\mathrm{eff})^2)}  \,.
\ee
\end{widetext}
This expression can be minimized with respect to the coherent feedback parameters analytically. Doing this yields the optimal values of $b=0$ and $\kappa_\mathrm{eff} = 2G$. Since $\kappa_\mathrm{eff} = 2\kappa(1-a)$ and the feedback parameter $a$ satisfies $a^2<1$, $\kappa_\mathrm{eff}$ can take values in the range $0<\kappa_\mathrm{eff}<4\kappa$. This means that the optimal cooling when $\kappa_\mathrm{eff} = 2G$ can be achieved for any $G<2\kappa$ by setting $a= 1-\frac{G}{\kappa}$. The optimal cooling is achievable for all weak couplings $G<\kappa$.  If this optimal setup is used, we obtain a steady-state mechanical covariance matrix with eigenvalue:
\be
    \sigma_m^{opt}=\frac{4 G^2 N_l + \Gamma_m (4 G + \Gamma_m) N_m}{(2 G + \Gamma_m)^2} \, .
\ee
Recall that steady state mechanical excitations are related to the mechanical eigenvalue through the expression $\Bar{N}  = (\sigma_m^{opt} -1)/2$. At this point we note that, though $\kappa_\mathrm{eff} =2G$ results in the optimal cooling, any feedback loop which brings $\kappa_\mathrm{eff}$ closer to the optimal value of $2G$ will improve the performance of the cooling. This can be seen by differentiating (\ref{sigcf}) with respect to $\kappa_\mathrm{eff}$ and setting $b=0$. This feature is also demonstrated in Figure \ref{sigmvskeff} which shows a plot of mechanical excitations against $\kappa_\mathrm{eff}$ with the minimum at $\kappa_\mathrm{eff}=2G$.

As well as decreasing steady state temperature of the system, CF can be used decrease the time taken for the system to relax. This is demonstrated in Figure \ref{TransientDyanmics} where the mechanical eigenvalue is plotted against time for systems with different CF setups. We find that, as the effective cavity loss rate is brought closer to the optimal value of $\kappa_\mathrm{eff}= 2G$, the rate of relaxation dramatically increases. This is because, when the mechanical loss rate $\Gamma_m$ is fixed, changing the relative value of $\kappa_\mathrm{eff}$ and $G$ results in large changes to the timescale on which the system operates. 

\subsubsection{Strong Coupling}
Now, we look at cooling with passive feedback in the strong coupling regime. By combining the Hamiltonian matrix of (\ref{HStrongxp}) with the CF modifications from (\ref{passivedrift}) we obtain the drift matrix
\be \label{SCdrift}
    A =
    \bmat
        -\frac{\kappa_\mathrm{eff}}{2} & - \Delta + 2\kappa b & 0 & 0\\ 
        \Delta -2\kappa b & -\frac{\kappa_\mathrm{eff}}{2} & -2G&0\\
        0 & 0 & -\frac{\Gamma_m}{2} & \omega_m\\
        -2G & 0 & -\omega_m & -\frac{\Gamma_m}{2}\\
    \emat \, .
\ee
The system will still have a diffusion matrix given by (\ref{passivediffusion}).
In the strong coupling regime, it is no longer possible to find a simple description of the optimal coherent feedback protocol analytically, and we must investigate this setup numerically. Nonetheless, there are some prelimilary observations we can make.

Recall that, in order for sideband driving to be effective, the finesse of the cavity must be high enough that the sidebands can be resolved. In practice, this means that the cavity loss rate must be much smaller than the mechanical frequency. For our purposes, we will take this to mean that we require $\kappa_\mathrm{eff}<0.1$. We note that coherent feedback allows a cavity with an otherwise low finesse to be brought into the sideband resolved regime through the lowering of $\kappa_\mathrm{eff}$. Conversely, by increasing $\kappa_\mathrm{eff}$, we can stabilise setups with large $G$-values that would otherwise be unstable. This is useful as, for large values of $G$ and small values of $\kappa_\mathrm{eff}$ the matrix (\ref{SCdrift}) is not always Hurwitz when the red sideband is driven. 

It turns out that the optimal $\kappa_\mathrm{eff}$ for cooling using the full linearized Hamiltonian lies in the range $G\lessapprox \kappa_\mathrm{eff}\lessapprox 2G$, depending on the coupling strength. Clearly, for couplings $G>0.05$, making $\kappa_\mathrm{eff}$ this high brings the system out of the resolved sideband regime, so this protocol cannot be used. Nonetheless, coherent feedback can be used to increase $\kappa_\mathrm{eff}$ as high as it can go without leaving the resolved sideband regime (i.e., $\kappa_\mathrm{eff}\simeq 0.1$). Doing this proves to be the optimal passive feedback protocol.

As an example of the efficacy of passive coherent feedback in the strong coupling regime, we investigate a setup with $\kappa = 0.025$, $\Gamma_m = 10^{-3}$, $N_m=200$, $N_l=1$ and a range of $G$ values. Figure \ref{StrongCouplingCooling} shows the minimum steady state mechanical excitations achievable by optimising the coherent feedback protocol for this setup in strong coupling regime, with the extra condition that $\kappa_\mathrm{eff}<0.1$. We find that in the strong coupling regime, passive feedback can still improve the performance of cooling.

\subsection{Active Feedback}
Now we investigate cooling using feedback loops involving squeezing and losses, as described in Section \ref{SqueezingCF}. In the weak coupling regime, if the red sideband is driven, the system evolves with a drift matrix given by
\be
    A = 
    \bmat
        \kappa(\frac{3}{2} \eta z - \frac{1}{2}\frac{\eta}{z} -1) & 0 & 0 & G \\
        0 & \kappa(\frac{3}{2}\frac{\eta}{z}-\frac{1}{2}  \eta z -1) & -G & 0\\
        0 & G & -\frac{\Gamma_m}{2} & 0 \\
        -G & 0 & 0 & -\frac{\Gamma_m}{2}
    \emat \, ,
\ee
and a diffusion matrix given in Sec.~\ref{SqueezingCF}. The steady state mechanical covariance matrix of this setup can be optimsed numerically with respect to the feedback paramters $\eta$ and $z$. We find that,  for any given setup in the weak coupling regime, the lowest entropy steady state is achieved when $z=1$ and $\eta = 1-\frac{G}{\kappa}$ which corresponds  to the optimal passive setup described in the previous section. In other words, adding squeezing in this way does not lead to better cooling of the mechanical oscillator. We conclude that interference alone is superior for CF-assisted cooling, without resorting to active operations. Thus, the 
addition of active operations, which are known to represent a resource 
in several contexts, does not seem to be useful for CF in this setting.

\subsection{Cooling with Delayed Feedback}
Until this point, we have assumed that any feedback was instantaneous. Here, we look at the effect of adding delays into the feedback loop, so as to obtain some understanding of the effect of delays on the optimal passive feedback loop described earlier. In particular, we will consider the case where the output of interface 1 is mixed at a beam splitter with an environmental mode $\rs_{in, l3}$ after a delay of $\tau$ before being immediately fed back into the cavity through interface 2. This amounts to setting
\bea
    \rs_{in l2}(t) = &a \rs_{out,l1}(t - \tau) + c \rs_{in, l3}(t)  \\ 
    = & a(\sqrt{\kappa} \rs_c(t-\tau) -\rs_{in,l1}(t-\tau)) + c \rs_{in,  l3}(t) \, , \nonumber 
\end{align}
which results in a delayed quantum Langevin equation for the system
\bea \label{delaylangevin}
    \dot{\rs}(t) =& A \rs(t) + a \kappa 
    \bmat
        \rs_c(t-\tau) \\
        0 \\
    \emat \\ 
    &+ 
    \bmat
        \sqrt{\kappa}(\rs_{in,l1}(t) - a \rs_{in,l1}(t-\tau) + c \rs_{in, l3}(t))\\
        \sqrt{\Gamma_m} \rs_{in,m}(t)
    \emat \, . \nonumber
\end{align}
Now, we define the Fourier transform of an operator as $\mathcal{F}[\hat{o}(t)]=\hat{o}(\omega) = \frac{1}{\sqrt{2 \pi}} \int_{-\infty}^{+\infty} \hat{o}(t)e^{i \omega t}dt$.  We note that $\mathcal{F}[\dot{\rs}] = - i \omega \rs(\omega)$ and $\mathcal{F}[\hat{o}(t-\tau)] = e^{i \omega \tau}\hat{o}(\omega)$. Fourier transforming equation (\ref{delaylangevin}) yields
\be \label{freqdomain}
    -i \omega \rs(\omega) = \Tilde{A}(\omega)\rs(\omega) + B(\omega) \rs_{in}(\omega) \, ,
\ee
where:
\be
    \Tilde{A}(\omega)=(A  + (\id_2 \oplus 0_2)a \kappa e^{i \omega \tau}) \, ,
\ee
\be
    B(\omega) =
    \bmat
        \sqrt{\kappa}(1-ae^{i\omega \tau}) \id_2 & \sqrt{\kappa}c \id_2 & 0 \\
        0 & 0 & \sqrt{\Gamma_m}\id_2
    \emat \, ,
\ee
\be
    \rs_{in}(\omega) =
    \bmat
        \rs_{in, l1}(\omega)\\
        \rs_{in, l3}(\omega)\\
        \rs_{in, m}(\omega)\\
    \emat \, .
\ee
We note that, as we have defined it, the Fourier transform of a Hermitian operator is not Hermitian, so in order to investigate physical observables, we must transform back into the time domain. This is done by re-arranging (\ref{freqdomain}) to get
\be
    \rs(\omega) = [-i\omega \id - \Tilde{A}(\omega)]^{-1} B \rs_{in}(\omega) = R(\omega)\rs_{in}(\omega) \, ,
\ee
where $R(\omega) =[-i\omega \id - \Tilde{A}(\omega)]^{-1} B $ is sometimes known as the transfer function of the system. We then  invert the Fourier transform to obtain the time-domain covariance matrix:
\be \label{sigt}
        \sig(t) = \frac{1}{2 \pi} \int_{-\infty}^{+\infty} d \omega d \omega' \langle \{\rs(\omega), \rs(\omega')^{\sf T} \}  \rangle e^{-i (\omega + \omega')t} \, .
\ee
The resulting covariance matrix will turn out not be be dependent on time and will in fact be the steady state covariance matrix for the system. The reason for this is that the only solution to our delayed differential equation for which the Fourier transform exists is the time-independent one. The Fourier transform is not defined for any time-dependent solutions to the differential equation. Therefore, when we Fourier transform the equation,  we are implicitly discarding all solutions except the steady state. 

By Fourier transforming the time domain input correlation functions (\ref{inputproperties1}), we obtain the the corresponding frequency domain relations
\bea
    [\rs_{in}(\omega), \rs_{in}(\omega')^{\sf T}] &= i \Omega \delta(\omega+ \omega') \; , \\
    \langle\{\rs_{in}(\omega), \rs_{in}(\omega')^{\sf T} \}\rangle &= \sig_{in} \delta(\omega+\omega') \; .
\end{align}
Plugging these relations into (\ref{sigt}) and integrating over the resulting delta functions yields the following expression for the steady-state covariance matrix of the system
\bea
    \sig(t) = \frac{1}{4 \pi} \int_{-\infty}^{+\infty} d \omega \bigg[& R(\omega)(\sig_{in}+ i \Omega)R(-\omega)^{\sf T}  \nonumber\\
   &+ R(-\omega) (\sig_{in} -i\Omega)R(\omega)^{\sf T}\bigg] \, .
\end{align}
This can be numerically evaluated. As an example, we use the setup described previously, where $\kappa = 0.1$,  $\Gamma_m = 10^{-5}$, $G=10^{-3}$, $N_l=1$, $ N_m =200$ and apply the optimal coherent feedback protocol by setting $a=1-\frac{G}{\kappa} = 0.99$. In the limit where $\tau=0$ (no delays) the average steady-state mechanical excitation number  is  $\Bar{N}^0= 0.988$. Delays of $\tau =1$, $\tau = 2$, and $\tau = 20$ result in steady state mechanical excitation numbers of 1.036, 1.084 and 1.939 respectively. From this, we can see that in-loop delays reduce the performance of the coherent feedback loops, but only to a small degree when the delays are on the order of the mechanical oscillation time period. In the limit of infinite delay $\tau \xrightarrow{} \infty$, since any modes being output from the cavity take an infinite amount of time to return, the system behaves as if no feedback loop is present.
\section{Optomechanical Entanglement} \label{EntanglementSection}

Now, we look at the ability of CF to improve the entanglement between the optical and mechanical oscillators. This is achieved by driving the blue sideband, i.e., $\Delta=\omega_m$. 
Here, we will confine our investigation to the weak coupling regime. 

Even in the weak coupling regime, driving the blue sideband often makes the system unstable, meaning that no steady state is reached. In this section, we will consider two tasks: stabilising unstable setups and increasing the entanglement of stable setups.
We quantify entanglement between the light and mechanics using the logarithmic negativity \cite{vidal2002computable}. For a bipartite Gaussian state with a covariance matrix of the form given in (\ref{generalCM}), it has a simple expression in terms of the matrix sub-blocks, see e.g. Ref.~\cite{serafini2017quantum}.
\subsection{Passive Coherent Feedback}
Recall that passive CF has the ability to tune the  effective cavity loss rate $\kappa_\mathrm{eff}$ and manipulate the optical cavity frequency.
Here, we focus on the tuning of $\kappa_\mathrm{eff}$, as a preliminary investigation suggested that tuning the cavity frequency was not useful in this context.

A system subject to blue sideband driving in the weak coupling regime with passive coherent feedback evolves according to the drift matrix
\be \label{weakbluepassive}
    A_{blue} =
     \bmat
        -\frac{\kappa_\mathrm{eff}}{2} \id_2 & -G \sigma_x \\
        -G \sigma_x & -\frac{\Gamma_m}{2} \id_2
     \emat \, ,
\ee
where $\sigma_x$ is the Pauli $x$-matrix and a diffusion matrix given by (\ref{passivediffusion}). For the  system to be stable, the drift matrix must be `Hurwitz', meaning that the real parts of its eigenvalues must all be negative. The eigenvalues of (\ref{weakbluepassive}) are given by
\be
    \lambda = \frac{1}{4}\big( -\Gamma_m - \kappa_\mathrm{eff} \pm \sqrt{16G^2 + \Gamma_m^2 - 2 \Gamma_m \kappa_\mathrm{eff}+ \kappa_\mathrm{eff}^2}\big) \, .
\ee
Thus, in order for the setup to be stable, we must have $\kappa_\mathrm{eff}>4G^2/\Gamma_m$. 
Through CF, the effective cavity loss rate can be tuned, to a maximum of $\kappa_\mathrm{eff} = 4 \kappa$. Therefore, CF can stabilise the blue sideband by increasing the effective cavity loss rate so that it satisfies $\kappa_\mathrm{eff}>4G^2/\Gamma_m$, provided that the unmodified parameters satisfy $\kappa>G^2/\Gamma_m$. However, for some setups with low initial $\kappa$, this might not be the case, so the system cannot be stabilized this way. Recall that, as described in Section \ref{TMSsection}, we can use two-mode squeezers to make the effective cavity loss rate arbitrarily high, at the cost of increasing the noise. We briefly investigate this in Section \ref{SectionEntanglement2mode}.

\begin{figure*}
    \centering
    \subfigure[]{\includegraphics[width=0.45\textwidth]{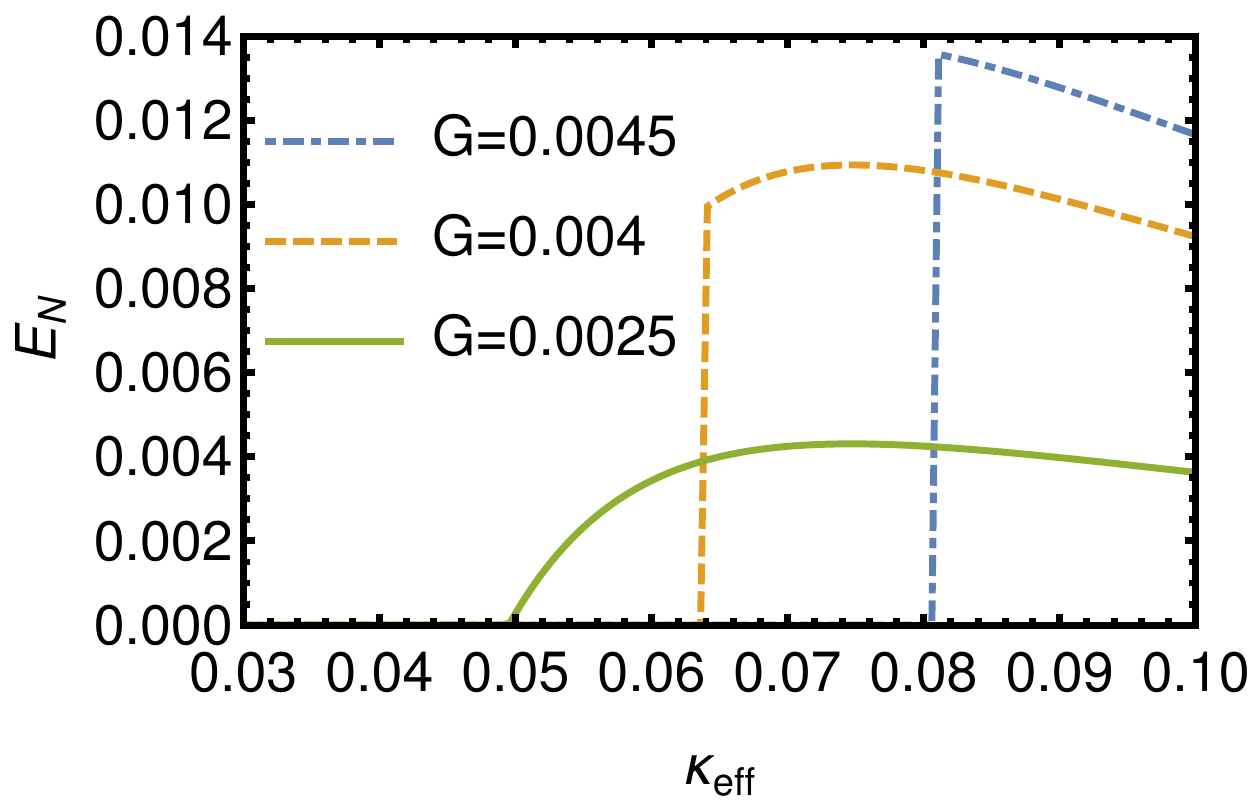} \label{PassiveBlueSideband}} 
    \hfill
    \subfigure[]{\includegraphics[width=0.45\textwidth]{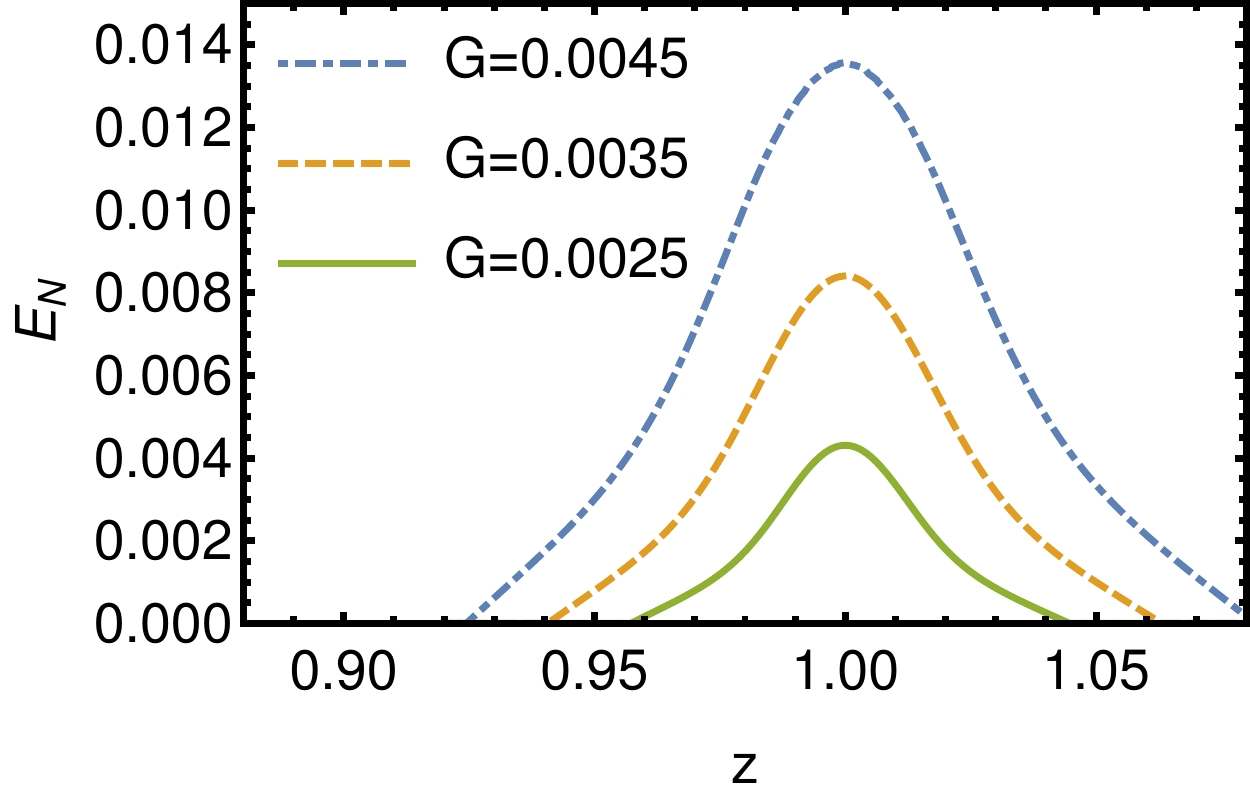} \label{BlueSidebandSqueezing}} 
    \caption{(a) The steady state logarithmic negativity against $\kappa_\mathrm{eff}$ for three setups, each with $\Gamma_m =10^{-3}$, $N_l=1$, $N_m=100$. The values of $\kappa_{eff}$, $G$ and $\Gamma_m$ are given in units where the mechanical frequency is equal to 1. The sudden vertical increases from $E_N$ seen for $G=4\x10^{-3}$ and $G=4.5\x 10^{-3}$ occur as the effective cavity loss rate becomes large enough to stabilise the system. (b) A plot of the maximum achievable logarithmic negativity $E_N$ against inloop squeezing $z$ for three setups with different coupling strengths $G$. The other parameters for each system are the same and have values $\kappa=0.1$,
    $\Gamma_m=10^{-3}$, $N_l=1$ and $
    N_m=100$.}
\end{figure*}

Now, we look at the ability of CF to increase the entanglement generated by a stable setup. 
As an example, we will consider a setup with $G=4.5\x10^{-3}$, $\Gamma_m=10^{-3}$ and $\kappa=0.1$. This setup is stable and, at steady state, the system has a logarithmic negativity of $E_N = 0.01166$. Numerically, we find that the stable logarithmic negativity optimized by making $\kappa_\mathrm{eff}$ as small as possible, without violating the stability criterion. This amounts to setting $\kappa_\mathrm{eff} = 4G^2/\Gamma_m +\epsilon \approx 0.081$, where $\epsilon$ is a small positive number required to maintain stability. Such a setup results in a stable logarithmic negativity of $E_N = 0.0138$. This increase is modest in relative terms but small in absolute terms. 
Figure \ref{PassiveBlueSideband} shows a plot of the stable logarithmic negativity against $\kappa_\mathrm{eff}$ for three setups, including the one described above. If the system is not stable, the $E_N$ is recorded as 0.  Interestingly, for the weaker couplings ($G=4 \x 10^{-3}$ and $G=2.5\x 10^{-3}$), the optimal protocol does not involve setting $\kappa_\mathrm{eff}$ to the minimum stable value and instead requires $\kappa_\mathrm{eff}$ to be higher than $4G^2/\Gamma_m$. Nonetheless, we find that for all three setups, tuning $\kappa_\mathrm{eff}$ can have some small but positive effect on $E_N$.

\subsection{Coherent Feedback with Squeezing and Losses}
Now we investigate the effect on entanglement generation of adding squeezing in-loop, as described in section \ref{SqueezingCF}. Since adding squeezing in-loop adds energy to the system, active CF will not be any better at stabilising unstable loops than passive CF. However, we can still investigate the efficacy of active CF for enhancing entanglement.

Fig. \ref{BlueSidebandSqueezing} shows the maximum steady state logarithmic negativity against in-loop squeezing. At each value of $z$, the logarithmic negativity has been optimized by tuning the beam splitter parameter $\eta$. We find that the logarithmic negativity of the system peaks when $z=1$, i.e., when there is no squeezing in-loop and the feedback loop reduces to the optimal passive loop considered in the previous section.
 
\subsection{Coherent Feedback with two-mode squeezing} \label{SectionEntanglement2mode}
Here, we consider loops containing two-mode squeezing, combined with phase shifters, whose evolution is described by the matrices in (\ref{TMSdriftdiffusion}). These loops have the property that the effective cavity loss rate can be increased arbitrarily by increasing the two mode squeezing. When loops of this kind are use in systems where the blue sideband is driven in the weak coupling regime, the drift and diffusion matrices read
\be
    A=
    \bmat
        -\frac{\kappa_S}{2}\id_2 & -G\sigma_x \\
        -G \sigma_x & -\frac{\Gamma_m}{2}\id_2
    \emat
    , \;
    D_S=
    \bmat
        \kappa_{S} N_S \id_2 & 0 \\
        0 & \Gamma_m N_m \id_2
    \emat ,
\ee
where $\kappa_S = 2 \kappa(1+\cosh{r})$, $N_S  = N_l\cosh{r}$ and $r$ is the two-mode squeezing parameter. Since increasing $r$ decreases the eigenvalues  of the drift matrix, loops of this kind can be used to stabilise systems. In particular, they are useful when $\kappa$ is too small for the system to be stabilized using passive CF. 

As an example, we consider a system with $\kappa = 0.01$,  $\Gamma_m =10^{-3}$,  $G = 4.5\x 10^{-3}$. We find that the blue sideband can be stabilized by increasing $r$ so that $\kappa_S>4G^2/\Gamma_m$,  which corresponds to $r>\cosh^{-1}(\frac{2G^2}{\kappa \Gamma_m}-1) \approx 1.78$.
However, though the system is stable with $r=1.78$, we find that the increase in noise (in the form of $N_S$) means that the logarithmic negativity of the steady state is zero. The ability of such loops to stabilise drastically unstable systems is potentially useful in other contexts, but shows fundamental limits in this context.
\section{Optical and Mechanical Squeezing}  \label{SqueezingSection}
We have already seen that a red sideband setup with passive feedback leads to thermal steady states. We now ask whether adding squeezing in the CF loop can be used to generate mechanical squeezing, which would be especially useful for sensing and metrology. We will therefore consider a setup in the red sideband regime,  subject to a coherent feedback loop of the form described in Section \ref{SqueezingCF}.

\begin{figure*}
    \centering
     \subfigure[]{\includegraphics[width=0.3\textwidth]{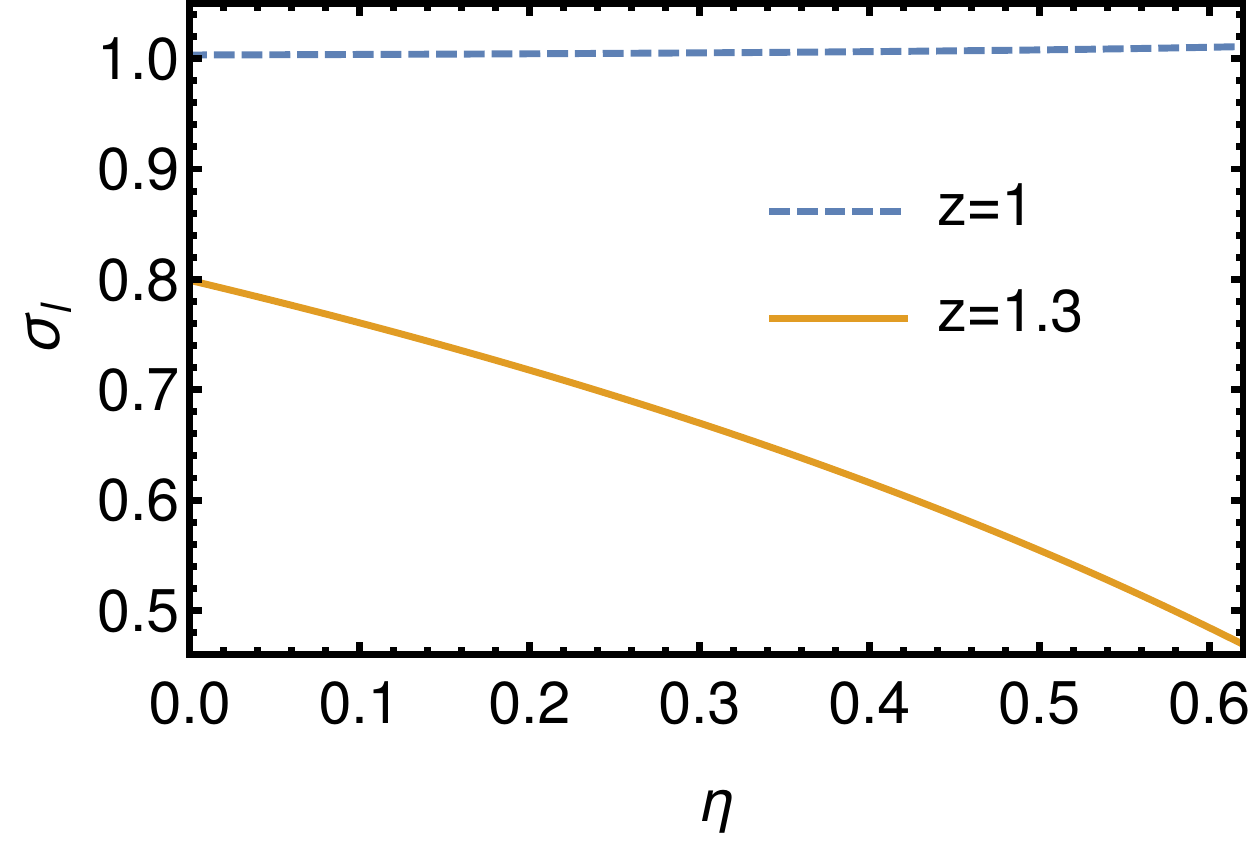} \label{opticalsqueezing}}
    \hfill
    \subfigure[]{\includegraphics[width=0.3\textwidth]{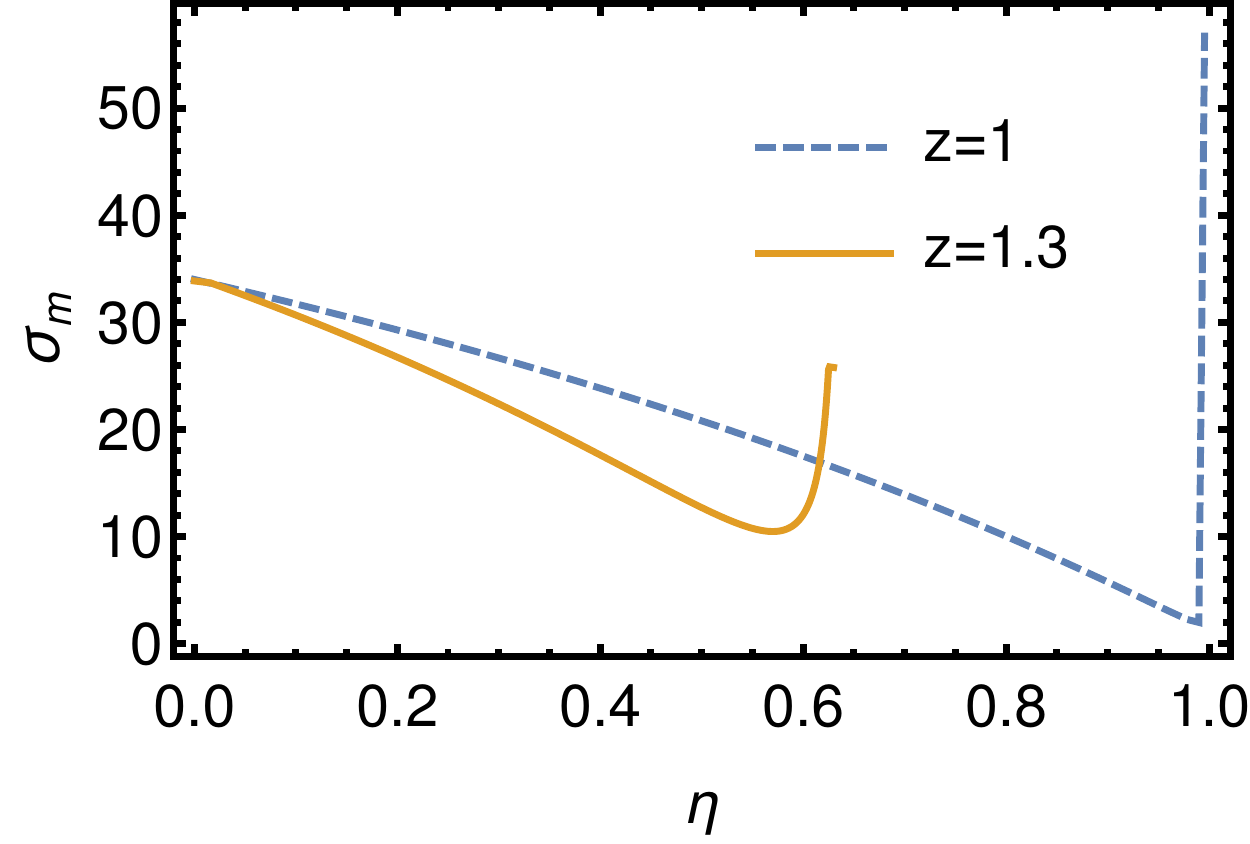} \label{mechanicalsqueezing}}
    \hfill
    \hfil\subfigure[]{\includegraphics[width=0.3\textwidth]{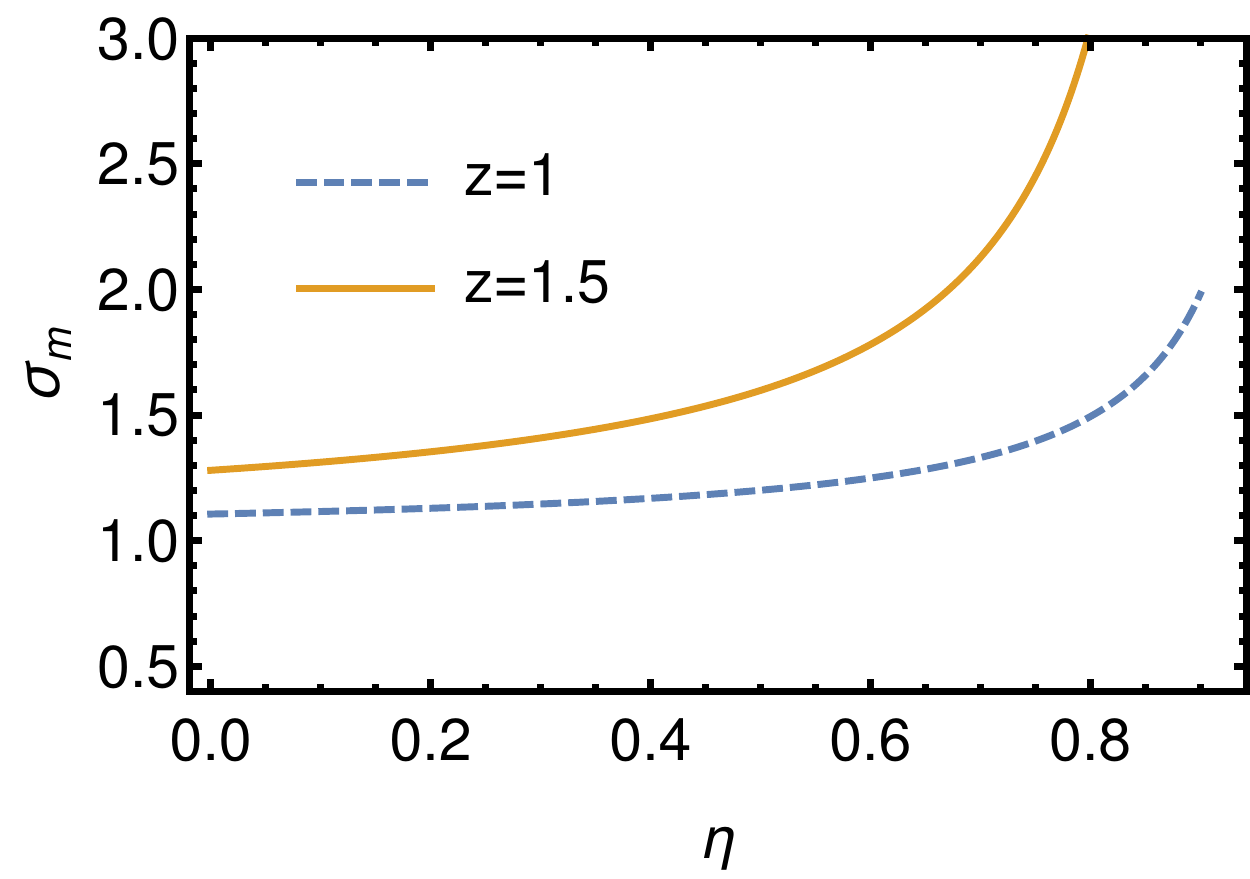} \label{mechanicalsqueezingStrong}}
    
    \caption{(a) The smallest steady state optical eigenvalue against beam splitter parameter $\eta$ for two different in-loop squeezings. The setup has $\kappa=0.1$, $G=10^{-3}$, $\Gamma_m = 10^{-5}$, $N_l=1$, $N_m = 100$.
    (b) The smallest steady state mechanical eigenvalue against beam splitter parameter $\eta$ for two different in-loop squeezings. The setup has $\kappa=0.1$, $G=10^{-3}$, $\Gamma_m = 10^{-5}$, $N_l=1$, $N_m = 100$. The values  for $z=1.3$ are only shown in the range $0<\eta<0.6388$ where the setup is stable. (c) The smallest steady state mechanical eigenvalue against beam splitter parameter $\eta$ for two different in-loop squeezings in the strong coupling regime. The setup has $\kappa=0.05$, $G=0.2$, $\Gamma_m = 10^{-4}$, $N_l=1$, $N_m = 100$. The setup with $z=1.5$ is stable for $\eta< 0.924$. }
\end{figure*}

The first thing we note is that such a setup, with $z\neq 1$, leads to stabilized squeezing of the {\em optical} mode. This can be seen in Figure \ref{opticalsqueezing}, which shows the smallest steady state optical eigenvalue against the beam splitter parameter $\eta$ for two setups, one passive with $z=1$ and one with an inloop squeezing of $z=1.3$. The active feedback loop is stable as long as $\eta<0.6388$. Note that this active feedback loop allows for stable squeezing of the optical mode below the vacuum noise for all values of $\eta$, and allows for optical squeezing with eigenvalue below $1/2$ (the so-called ``3dB limit'') for $\eta>0.59$. 

Figure \ref{mechanicalsqueezing} shows the smallest steady {\em mechanical} eigenvalues for both the squeezed and passive feedback loops in the same setup. We can see that adding in-loop squeezing can reduce the noise on a mechanical quadrature, but cannot outperform the optimal passive loop for cooling, which is achieved by setting $\eta = 1-\frac{G}{\kappa}$ with $z=1$. 

We also investigate the same loop in the strong coupling regime using the full Hamiltonian  (\ref{HStrongxp}) with $\Delta = -\omega_m$ and find again that the passive loop outperforms the active loop. This is demonstrated in Figure \ref{mechanicalsqueezingStrong}. 

Rather surprisingly, we could identify no CF loop that would push the noise of a mechanical quadrature below 
the vacuum level. Our investigation therefore indicates that CF is not useful for the generation of mechanical squeezing. This stands in contrast with measurement-based protocols, which can be extremely effective for generating mechanical squeezing (although conditionally, in the absence of feedback)~\cite{Vanner11,CMJ08,Genoni15,Me19,Me20}.
\section{State  Transfer} \label{TransferSection}

Another promising application of optomechanical systems is as transducers converting between optical and mechanical states.
Therefore, in this section we look at CF as a method for enabling and assisting the transfer of a state from the optical mode to the mechanical mode. This is facilitated by the red sideband interaction. In particular, we aim to prepare the mechanical mode in a state with a `target' covariance matrix $\sig_T$. For state transfer we will consider a setup where the light mode is initialized in the target state with covariance matrix $\sig_l(0)=\sig_T$, and the mechanical mode is initialized in the thermal state with $\sig_m(0) = N_m \id_2$.
If left alone for an indefinite amount of time, the system will revert to a steady state which has no relation to the initial state. Therefore, to quantify state transfer, we will need to investigate the transient dynamics of the system. The efficacy of the state transfer at time $t$ will be quantified by the function $V(t)=|| \sig_T -\sig_m(t)||_2$ where $||\cdot||_2$ indicates the Schatten 2-norm given by $||M||_2 = \sqrt{\text{Tr}[M^2]}$.
Our figure of merit for this study will be the minimum value of $|| \sig_T -\sig_m(t)||_2$ reached during a period of 25 mechanical oscillations for which the red sideband is pulsed. In order to verify that the mechanical state is indeed a result of state transfer, we compare our results to a setup initialized with the optics in a thermal state $\sig_l = N_l \id_2$. If the minimum value of $|| \sig_T -\sig_m(t)||_2$ is smaller when we start the optics in state $\sig_T$, we can argue that the improvement must have come from state transfer.

\subsection{The Strong Coupling Regime}
Since preliminary results suggest that state transfer is poor in the weak coupling regime, we will investigate state transfer in the strong coupling regime. In particular, we will investigate the effect of tuning the effective cavity loss rate through passive coherent feedback. The covariance matrices of all single-mode pure Gaussian states be written $\sig = R_{\theta} Z R_{\theta}^{\sf T}$ where $Z= \text{diag}(1/z, z)$ gives the squeezing of the state, and $R_{\theta}$ is a $2 \x 2$ rotation matrix with angle $\theta$. Thus, all Gaussian single-mode pure covariance matrices can be parameterized in terms of the two values $\theta$ and $z$. Since the full red sideband Hamiltonian, given by equation (\ref{HStrongxp}) with $\Delta=-\omega_m$ is phase-dependent, we will expect the efficacy of state transfer to depend on the rotation angle $\theta$. As a result, when investigating state transfer,  we choose a state with fixed squeezing $z=1/4$ and average over the possible rotation angles $\theta$. 

\begin{figure*}
    \centering
     \subfigure[]{\includegraphics[width=0.45\textwidth]{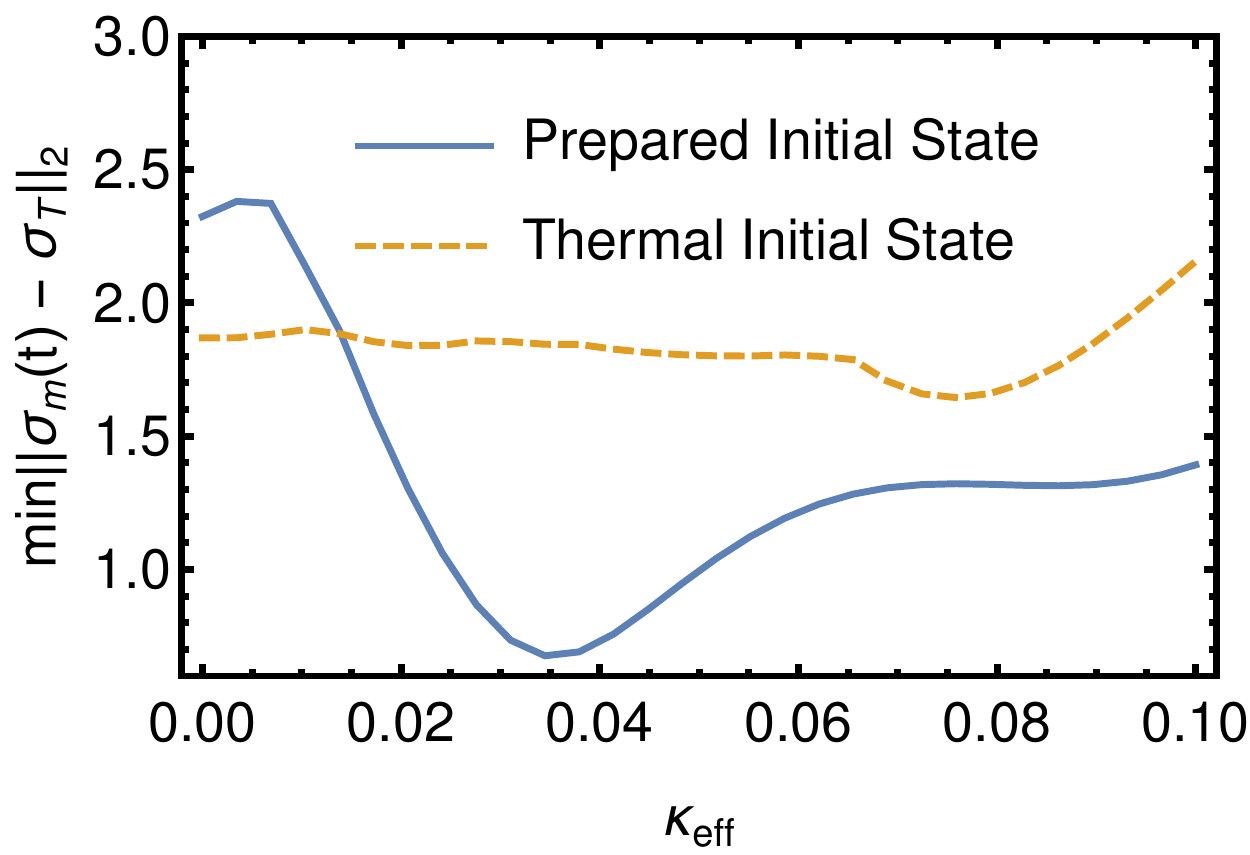} \label{STfig}}
    \hfill
    \subfigure[]{\includegraphics[width=0.45\textwidth]{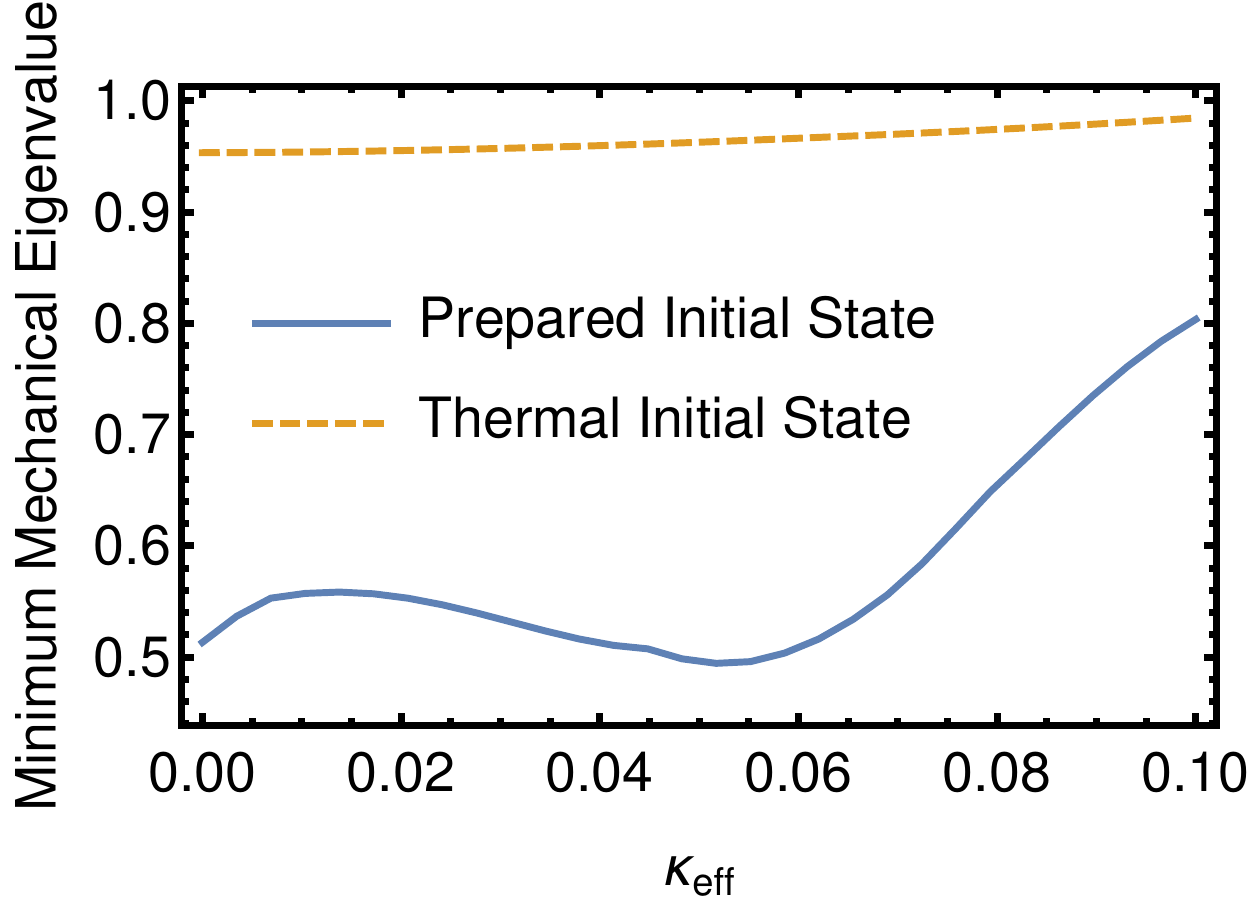} \label{SqueezingTransfeFig}}
    
    \caption{(a) The efficacy of state transfer from optics to mechanics for a range of $\kappa_\mathrm{eff}$ values. The setup considered had $G=0.1$, $N_l=1$, $N_m=100$, $\Gamma_m = 10^{-5}$. The values of $\kappa_{eff}$, $\Gamma_m$ and $G$ are all given in units where the mechanical frequency is equal to 1. The state considered for transfer was a squeezed state with $z=1/4$ and data was averaged over the rotation angle $\theta$ of the squeezed state. (b) The minimum mechanical eigenvalue reached during the first 25 mechanical oscillations, against the effective cavity loss rate $\kappa_\mathrm{eff}$ (in units where the mechanical frequency is equal to 1). The setup uses $G=0.1$, $N_l=1$, $N_m=100$, $\Gamma_m = 10^{-5}$.}
\end{figure*}

Figure \ref{STfig} shows the minimum value of $V(t)$ achieved within the first 25 mechanical  oscillations which occurs when the red sideband is pulsed. Results shown in blue (with a solid line) were generated by starting the optical mode in the desired state $\sig_T$ and recording the closest approach to this state achieved by the mechanical mode. The results in orange (displayed with a dashed line) were collected in a similar way, but the optics were initialized in a thermal state. Therefore, when the blue (solid) line is lower than the orange (dashed) line, we can say that preparing the optical mode in a desired state and performing  state transfer is more effective at achieving a certain mechanical state than normal evolution. 

We can now appreciate the effect of tuning the cavity loss rate through coherent feedback on boosting the performance of state transfer, illustrated in Figure \ref{STfig}. Interestingly, performance does not always increase with lower $\kappa_\mathrm{eff}$,  and is instead optimized at  $\kappa_\mathrm{eff} \approx 0.035$. We remind the reader that cavity loss rate is normally a fixed parameter of the system and the ability to tune it to optimize state transfer is provided by coherent feedback.

We may ask whether the states transferred to the mechanical mode share the salient features of  the prepared state. In particular, we can look at whether the state transfer protocol described above is effective at transferring squeezing to the mechanical oscillator. Figure \ref{SqueezingTransfeFig}  shows the minimum mechanical eigenvalue recorded during the first 25 mechanical oscillations for a thermal inital state and prepared optical initial states. As before, the mechanical oscillator is assumed to be initialized in a thermal state, and the optical mode is initialized in a squeezed state with $z=1/4$. Again, the results are averaged over the rotation angle of the prepared state $\theta$. The figure demonstrates that preparing the optical mode in a squeezed state leads to greater transient mechanical squeezing than a thermal state, and that the amount of the mechanical squeezing can be altered by tuning the $\kappa_\mathrm{eff}$ through coherent feedback. We note that the coherent feedback loops in Section \ref{SqueezingSection} provide a possible method for preparing an initial optical squeezed state.

\section{Conclusion}
We have investigated the ability of coherent feedback to improve the performance of several tasks in a linearized optomechanical setting. For the tasks of optimising steady state cooling and entanglement,  as well as transient state transfer, we have found that coherent feedback can improve performance of the system when measured with an appropriate figure of merit. However, with the exception of cooling, the improvements yielded are modest, and this study provides insight into the limitations of coherent feedback. Though coherent feedback allows for a wide range of manipulations to the system (as detailed in Section \ref{CFSection}), in practice, most of the benefits found in this study derive from the ability to tune the effective cavity loss rate through passive coherent feedback. Indeed, for most tasks, adding active elements to the feedback loop is detrimental to the performance of the setup. A plausible explanation for this is that active elements, by definition, add energy to the system, adding noise and destabilising the setup. As a result, the benefits of active feedback (in the examples we considered) are often outweighed by these drawbacks.

We should note that, aside from the investigation of state transfer, we focused on steady-state features, and that coherent feedback techniques would not be constrained by the same limitation if, instead, one were to consider transient dynamics. This could be an interesting direction for future work building up on the general formalism presented in this study. Since coherent feedback allows features of the cavity (such as the loss rate and the frequency) to be easily tuned through passive elements in the feedback loop, one avenue for further research would be to investigate the time-dependent modulation, through coherent feedback, of these parameters which are normally assumed to be fixed. Dynamical modulation of the cavity dissipation rate has already been found to be beneficial for cooling the mechanical oscillator in an optomechanical setup \cite{dynamicCooling}. Such modulation could be achieved by  tuning the free carrier plasma density \cite{xu2007breaking, kondo2013ultrafast, soref1987electrooptical} or using light absorbers or scatterers in deformable optical cavities \cite{favero2009optomechanics}. We  argue that the method presented here using coherent feedback, being based entirely on optical elements in interferometric setups, may prove easier to implement than these proposals.

Another limitation of the investigation presented here is that all work takes place in the Gaussian regime, with linear dynamics. Lifting this restriction would allow inquiry into a much larger class of both physical phenomena and types of coherent feedback loop. 

\section{Acknowledgements}
M. B. acknowledges support by the European Union Horizon 2020 research and innovation programme under grant agreement No 732894 (FET Proactive HOT).
\bibliographystyle{apsrev4-1}
\bibliography{OptomechanicsCF}
\end{document}